# *PT* and Anti-*PT* Symmetry Arising in Time Dependent Weak Value Measurements

A. D. Parks, J. E. Gray*, V. H. Gehman Jr., and G. K Josemans

ℏ Laboratory, Electromagnetic and Sensor Systems Department, 18444 Frontage Road Suite 327, Naval Surface Warfare Center, Dahlgren Division, Dahlgren, VA 22448-5161, USA

*Corresponding author; email: john.e.gray@ieee.org

ORCID ID: 0000-0002-9260-9926

**Abstract:** Parks introduced a formulation of time dependent weak values in 2008, which is the formalism we use in this paper. In this paper we extend notions from time dependent weak values to show that Hamiltonians associated with weak value measurements can be shown to exhibit even or odd symmetric properties. They exhibit **PT** or anti-**PT** symmetry, respectively. These symmetries are manifested during the measurement process as pointer translations, which have vanishing imaginary or vanishing real parts. The consequence of this that one can characterize some of the aspects of these symmetries of the time dependent Hamiltonians that arise from time dependent weak values. This allows one to generalize some work on non-Hermitian variables in quantum mechanics (QM) due to Bender related to weak values and weak measurement. We also speculate how these symmetries might apply to distinguishing between the various two-time interpretations of QM in the Conclusions of this paper.

## 1 Introduction

The theoretical notion of the weak value of a quantum mechanical observable was introduced by Aharonov *et al* [1-3] more than three decades ago. A weak value is the statistical result of a standard measurement procedure performed upon a pre- and post-selected (PPS) ensemble of quantum systems when the interaction between the measurement pointer and each system is sufficiently weak. Unlike the standard strong measurement of a quantum mechanical observable, which significantly disturbs the measured system, a weak measurement of an observable for a PPS system does not appreciably disturb the quantum system and yields the complex valued weak value as the measured value of the observable. Although the interpretation of the theory has been somewhat controversial, a series of experiments performed in recent years has verified aspects of weak value theory [4], and the theory has been applied to such diverse areas as contextuality [5], retro-causality [6], Hardy's paradox [7], quantum computing [8], interferometry [9], sensing [10], and metrology [11].

Since the introduction of *PT* (i.e., parity-time reversal) symmetric quantum mechanical non-Hermitian Hamiltonians by Bender *et al* [12, 13] more than twenty years ago, their theoretical implications have been controversial and the subject of intense basic and applied research. Canonical quantum mechanics requires a quantum system's Hamiltonian $\widehat{H}$ obey the mathematical Dirac Hermitian symmetry condition $\widehat{H} = \widehat{H}^\dagger$, which implies the eigenvalues of $\widehat{H}$ are real valued, the associated time evolution is unitary, and probabilities are conserved. Such Hermitian systems describe the properties of idealized isolated systems. Non-Hermitian systems, i.e., those for which $\widehat{H} \neq \widehat{H}^\dagger$, have complex valued eigenvalues, are not unitary, and provide approximate descriptions of dissipative processes. The class of non-Hermitian Hamiltonians discovered by Bender *et al* are those that are invariant under application of parity *P* and time reversal *T* transformations and are intermediate between Hermitian and non-Hermitian Hamiltonians. For example, like Hermitian Hamiltonians, *PT* symmetric Hamiltonians have real valued eigen-spectra and, like non-Hermitian Hamiltonians, they can describe non-isolated systems. Since its discovery, *PT* symmetric systems have been confirmed to occur in a variety of settings [14–17], and has served as the basis for new devices [18], and novel metamaterials [19] as well.

Another interesting connection between weak values and non-Hermitian operators was recently published by Pati *et al* [20]. There an experimentally verifiable protocol was proposed for determining the complex valued mean value of any non-Hermitian operator from a direct measurement of the weak value of the operator's positive-semi-definite part. They also briefly examined the utility of their protocol for measuring $\mathcal{PT}$ symmetric non-Hermitian Hamiltonians.

There has been much interest in the philosophical implications of the two-time formalism of weak measurement as found in discussions by Aharonov and others [21-23]. Since this is not the focus of this paper, it will not be discussed here, other than to note [24]. While time independent weak values are widely discussed, time dependent weak value literature exists as well [25-28]. This paper introduces another connection between time dependent weak values and $\mathcal{PT}$ symmetric properties of non-Hermitian Hamiltonians. In particular, it is shown that there are non-Hermitian Hamiltonians associated with weak value measurement pointer translations that exhibit both $\mathcal{PT}$ and anti-$\mathcal{PT}$ symmetries when the measured weak values are both even and odd symmetric in time. Unlike the class of $\mathcal{PT}$ symmetric discovered by Bender et al, the eigen-spectra of such non-Hermitian Hamiltonians can be both continuous and complex valued; thus, the relationship between these Hamiltonians and the time symmetric properties of their weak values can be examined. Note some applications of time dependent weak values are also discussed relative to nuclear decay related topics [29-30], which is an interested application of them.

We address the following issues in each section. In **Section 2** that, unlike the class of $\mathcal{PT}$ symmetric Hamiltonians discovered by Bender *et al*, we are able to extend the eigen-spectra of these Hamiltonians to be both continuous and complex valued. The relationship between these Hamiltonians and the time symmetric properties of their weak values are identified. In **Section 3**, PT and Anti-PT Symmetric Weak Value Measurements, we discover that thus, the $\mathcal{PT}$ and anti-$\mathcal{PT}$ symmetry associated with weak value measurements of an observable depend strictly upon the even and odd time symmetries of the time dependent weak value $A_w(t)$. In **Section 4**, we examine the weak position and weak momentum pointer translations generated by $\mathcal{PT}$ and anti-$\mathcal{PT}$ symmetric weak value measurements when the coupling strengths are proportional to a function that is even symmetric in time around a time $t_0$. In **Section 5**, **Application to Arbitrary Pointer Observables**, theorems are directly applied to the recently reported properties of weak value measurement pointers for the general case where $\langle q|\varphi\rangle$ and $A_w$ are complex valued and the interaction Hamiltonian. In the **Conclusion**, we speculate about some additional connections to a paper by Dressel as well further directions for time dependent weak values.

## 2 The Dynamics of Weak Value Measurements

We assume the reader is familiar with the standard formulation of weak values that is outlined in the references presented in the introduction, but is not familiar with time dependent formulation that is found in Parks [25], so we review it in some detail in order to clarify future applications beyond what has been previously discussed in the literature. The derivation of time dependent weak values is found in Section 3 & 4. After the review we discuss various consequences of time dependent weak values that were not discussed in [25].

The key to understanding time dependent weak values is based on the observation in [25] which states, "*Fundamental to the theory of weak values is the proposition that although $\hat{A}$ occurs at $t_0$, the PPS are pre-selected at different times $t_i < t_0$ and $t_f > t_0$, respectively.* [Note while measurements ideally can be formulated in terms of Dirac delta distributions $\delta(t)$, this is an idealization. Measurement at a time $t_m$ is actually accomplished in finite time intervals $t_m \in [t_i, t_f]$, this true both in classical and quantum measurement.*] PPS states are selected at these times define past and future boundary conditions which influence $A_w$ at measurement time $t_0$ via their unitary evolution forward in a time frame from $t_i$ to $t_0$ and backward in time from $t_f$ to $t_0$. Such unitary evolutions are responsible for the weak energy of evolution at the time of measurement.*"

With time dependent weak values, we are stepping into the realm of time ordered operators when we try to understand the mechanism that underlies them. We can define $\mathcal{A}(t)$ as the time-ordered operator in the interval $\mathcal{T} \in [t_i, t_f]$, as
$$\mathcal{A}(t) \equiv \{A_w(t): t \in \mathcal{T}\}.$$
where $A_w(t)$ is the theoretical weak value of $\hat{A}$ at the interaction time t and T is a fixed closed time interval such that it has the two properties:

1. at each time $t \epsilon T$, the weak value $A_w(t)$ is defined in terms of the state $|\psi_i(t_i)\rangle$ which is pre-selected at the time $t_i = t - \Delta t_i$, and by a state $|\psi_f(t_f)\rangle$ which is post-selected at $t_f = t - \Delta t_f$, where both $\Delta t_i$ and $\Delta t_f$ are fixed time intervals;

2. The PPS states change continuously from their initial states at times $t_i = t - \Delta t_i$ and $t_f = t - \Delta t_f$ respectively via the Schrodinger equation

$$\frac{d|\psi_i(t_i)\rangle}{dt_i} = \frac{i}{\hbar}\hat{H}_i|\psi_i(t_i)\rangle, \tag{1a}$$

where $t_i \epsilon \in [t_1 - \Delta t_i, t_2 - \Delta t_i]$ as well as

$$\frac{d|\psi_f(t_f)\rangle}{dt_f} = \frac{i}{\hbar}\hat{H}_f|\psi_f(t_f)\rangle, \tag{1b}$$

where $t_f \in [t_1 - \Delta t_f, t_2 - \Delta t_f]$. Note that $\mathcal{A}(t)$ is the theoretical pointer profile for $A_w(t)$ over T, the set

$$\mathcal{A}_q(t) \equiv \{Re\ A_w(t): t \in \mathcal{T}\}$$

is the theoretical pointer position profile of the weak value observation, and the set

$$\mathcal{A}_p(t) \equiv \{Im\ A_w(t): t \in \mathcal{T}\}$$

is the theoretical profile of values which effect the pointer momenta distribution.

The usual approach to weak measurement and weak values tacitly assumes no time dependence, so the Hamiltonians are fixed in time, but this does not have to be the case. Instead, if $\hat{H}_i$ and $\hat{H}_f$ are functions of time, the time evolution is governed by the equations

$$\frac{d|\psi_i(t_i)\rangle}{dt_i} = exp\left(-\frac{i}{\hbar}\hat{H}_i\Delta t_i\right)|\psi_i(t_i)\rangle = \hat{U}_\iota\ |\psi_i(t_i)\rangle, \tag{2a}$$

and

$$\frac{d|\psi_f(t_f)\rangle}{dt_f} = exp\left(\frac{i}{\hbar}\hat{H}_f\Delta t_f\right)|\psi_f(t_f)\rangle = \hat{U}_f|\psi_f(t_f)\rangle. \tag{2b}$$

Then applying these to the weak value formula gives

$$A_w(t) = \frac{\langle\psi_f(t_f)|\hat{U}_f^\dagger\ \hat{A}\ \hat{U}_\iota|\psi_i(t_i)\rangle}{\langle\psi_f(t_f)\hat{U}_f^\dagger|\hat{U}_\iota\psi_i(t_i)\rangle} = \frac{\langle\psi_f(t)|\hat{A}|\psi_i(t)\rangle}{\langle\psi_f(t)|\psi_i(t)\rangle}. \tag{3}$$

The reason we can replace the indexed times with t is because when (2a) and (2b) govern time throughout out the time interval, we can replace the indexed terms with just $t$, where t means any time in the interval T. This replacement is possible because $[\hat{U}_\iota, \hat{H}_i] = 0 = [\hat{U}_f, \hat{H}_f]$.

Thus, the actions of $\hat{H}_i$ and $\hat{H}_f$ upon the associated PPS states at times $t_i$ and $t_f$ are transformed by the operators $\hat{U}$ and $\hat{V}$ into actions of the Hamiltonians operators upon the evolved PPS states at measurement time t. Equations (2a-b) become

$$\frac{d|\psi_i(t_i)\rangle}{dt} = -\frac{i}{\hbar}\hat{H}_i|\psi_i(t)\rangle, \tag{4a}$$

and

$$\frac{d|\psi_f(t_f)\rangle}{dt} = -\frac{i}{\hbar}\hat{H}_f|\psi_f(t)\rangle. \tag{4b}$$

since $\hat{U}$ and $\hat{V}$ are constant operators. If $\dot{A}_w(t)$ exists for each $t \in T$, then is a continuous function the interval T. Then the time derivative of $A_w(t)$ can be shown to be

$$\dot{A}_w(t) = \frac{i}{\hbar}\left[\left(H_f(t)A(t) - A(t)H_i(t)\right)_w - \left(H_f(t) - H_i(t)\right)_w\right] \tag{5a}$$

so, the t in the equation is the measurement time defined in terms of the t-dependent PPS states. The second term in (5) is the weak energy for the PPS which was first derived in [31] for time independent weak values. Note the actions for evolving forward and backward define the range of the time interval T. Note if both $t_i$ and $t_f$ are contemporaneous, then $\hat{H}_i = \hat{H}_f = \hat{H}$ and therefore

$$\dot{A}_w = \langle \dot{A} \rangle = \frac{i}{\hbar}[\hat{H}, \hat{A}]. \tag{5b}$$

(5b) tells us the present disappears if they are contemporaneous and time evolution is simply a continuous group. A less speculative of the time dependence of $A_w$ is that if defines phases and phase factors that play a role when the time dependence is explicit, $A_w(t)$.

As we have already noted, a weak measurement of an observable can in the form given is a function of time. A specific form for an observable is termed a measurement Hamiltonian $\hat{H}(t)$. So, after the measurement of an observable $\hat{A}$ of a PPS system, the pointer state is given in the specific form of the measurement Hamiltonian:

$$|\Phi\rangle = \left\langle \psi_f \middle| e^{-\frac{i}{\hbar}\int \hat{H}(t)dt} \middle| \psi_i \right\rangle |\varphi\rangle, \tag{6a}$$

where $|\psi_i\rangle$ and $|\psi_f\rangle$ are the PPS states, respectively, $|\varphi\rangle$ is the initial pointer state, and the measurement Hamiltonian can be expressed in either the momenta picture:

$$\hat{H}(t) = \gamma_p(t)\hat{A}\hat{p}, \tag{6b}$$

or the position picture:

$$\hat{H}(t) = -\gamma_q(t)\hat{A}\hat{q}. \tag{6c}$$

The first describes the coupling between the observable and the pointer momentum $\hat{p}$ while the second describes the observable's coupling to the pointer position $\hat{q}$, respectively. Here, either of the two possible $\gamma$ are the associated real-valued coupling strengths. When the measurement is a weak measurement and the PPS states are considered time-dependent, then the usual weak value approximation to (6a) yields

$$|\Phi(t)\rangle \approx \langle \psi_f(t)|\psi_i(t)\rangle e^{-\frac{i}{\hbar}\int \hat{H}_w(t)dt}|\varphi\rangle, \tag{7}$$

where we have $\hat{H}_w(t)$ becomes either the momenta interaction weak Hamiltonian

$$\hat{H}_{w,p}(t) = \gamma_p(t)A_w(t)\hat{p}, \tag{8}$$

or position interaction weak Hamiltonian

$$\hat{H}_{w,q}(t) = -\gamma_q(t)A_w(t)\hat{q}. \tag{9}$$

Note, that both Hamiltonians are the non-Hermitian *weak interaction Hamiltonians*, and $A_w(t)$ is the complex valued weak value of $A$ or Eq. (3) using the Von Neuman model Aharonov has used, so we get the same result:

$$A_w(t) = \frac{\langle \psi_f(t)|\hat{A}|\psi_i(t)\rangle}{\langle \psi_f(t)|\psi_i(t)\rangle}.$$

Application of Hamilton's equations to (8) and (9) yields

$$\frac{\partial \hat{H}_{w,p}(t)}{\partial \hat{p}} = \gamma_p(t)A_w(t) \equiv \dot{q}_w(t)$$

and

$$\frac{\partial \hat{H}_{w,q}(t)}{\partial \hat{q}} = -\gamma_q(t)A_w(t) \equiv -\dot{p}_w(t)$$

as the complex valued rates of change of the *weak pointer translations* during the measurement process. Equations (8) and (9) can then be written as

$$\hat{H}_{w,p}(t) \equiv \dot{q}_w(t)\hat{p} \tag{10}$$

and

$$\hat{H}_{w,q}(t) \equiv -\dot{p}_w(t)\hat{q}. \tag{11}$$

Using the fact that $[\hat{q}, \hat{p}] = i\hbar$, it is easy to show that the equations of motion for the *weak position* $q_w$ and the *weak momentum* $p_w$ translations can be written compactly in terms of commutators as

$$\dot{q}_w(t) = -\frac{i}{\hbar}[\hat{q}, \hat{H}_{w,p}(t)]$$

and

$$\dot{p}_w(t) = -\frac{i}{\hbar}[\hat{p}, \hat{H}_{w,q}(t)].$$

Also, from (6a), (8), and (9), it is clear the measurement produces weak position and weak momentum pointer translations given by

$$q_w = \int \dot{q}_w(t)dt = \int \gamma_q(t) A_w(t) dt$$

and

$$p_w = \int \dot{p}_w(t)dt = \int \gamma_p(t) A_w(t) dt.$$

Before leaving this section, it is noted the eigen-value spectra of $\hat{H}_{w,p}(t)$ and $\hat{H}_{w,q}(t)$ must necessarily be continuous and complex valued. This can be seen by using the identities

$$\langle q|\hat{H}_{w,p}(t)|\varphi\rangle = -i\hbar \dot{q}_w \frac{\partial \langle q|\varphi\rangle}{\partial q},$$
$$\langle p|\hat{H}_{w,p}(t)|\varphi\rangle = \dot{q}_w p \langle p|\varphi\rangle,$$
$$\langle q|\hat{H}_{w,q}(t)|\varphi\rangle = -\dot{p}_w q \langle q|\varphi\rangle,$$

and

$$\langle p|\hat{H}_{w,q}(t)|\varphi\rangle = -i\hbar \dot{p}_w \frac{\partial \langle p|\varphi\rangle}{\partial p}$$

to show that the actions of $\hat{H}_{w,p}(t)$ and $\hat{H}_{w,q}(t)$ upon the pointer state $|\varphi\rangle$ yield the eigen-value equations

$$H_{w,x}(t)\varphi(q) = \mathcal{E}_{x,q}(t)\varphi(q)$$

and

$$H_{w,x}(t)\bar{\varphi}(p) = \mathcal{E}_{x,p}(t)\bar{\varphi}(p),$$

where $x = p, q$, and $\bar{\varphi}(p)$ is the Fourier transform of $\varphi(q) = \langle q|\varphi\rangle = (2\pi\hbar)^{-\frac{1}{2}} e^{\frac{i}{\hbar}pq}$. The eigenvalues obtained from these equations are $\mathcal{E}_{p,q}(t) = \mathcal{E}_{p,p}(t) = \dot{q}_w(t)p$ and $\mathcal{E}_{q,p}(t) = \mathcal{E}_{q,q}(t) = -\dot{p}_w(t)q$ and are clearly continuous and complex valued.

## 3 *PT* and Anti-*PT* Symmetric Weak Value Measurements

Note that since $A_w(t)$ is complex valued, then both $\hat{H}_{w,p}(t)$ and $\hat{H}_{w,q}(t)$ are non-Hermitian operators. Consequently, they do not obey the Dirac symmetry condition. However, it is interesting to consider the behavior of these operators under the combined actions of the parity and time reversal transformations $\mathcal{P}: (\hat{q}, \hat{p}) \to (-\hat{q}, -\hat{p})$ and $\mathcal{T}: (\hat{q}, \hat{p}, t, i) \to (-\hat{q}, -\hat{p}, -t, -i)$, respectively, as defined in [29].

In what follows, *it will always be assumed that the coupling strengths $\gamma(t)$ and $\theta(t)$ are even symmetric in time*, i.e., $\gamma(t) = \gamma(-t)$ and $\theta(t) = \theta(-t)$. In this case, application of the $\mathcal{T}$ transformation followed by the $\mathcal{P}$ transformation to $\hat{H}_{w,p}(t)$ and $\hat{H}_{w,q}(t)$ yields

$$\hat{H}_{w,p}(t) \to \gamma(-t) A_w(-t)^* (-\hat{p}) \to \gamma(t) A_w(-t)^* \hat{p} = \hat{H}_{w,p}^{PT}(t)$$

and

$$\hat{H}_{w,q}(t) \to -\theta(-t) A_w(-t)^* (-\hat{q}) \to -\theta(t) A_w(-t)^* \hat{q} = \hat{H}_{w,q}^{PT}(t)$$

as the *PT* transformed operators $\hat{H}_{w,p}^{PT}(t)$ and $\hat{H}_{w,q}^{PT}(t)$. Here, *PT symmetric weak value measurements* are defined as those for which $\hat{H}_{w,p}(t)$ or $\hat{H}_{w,q}(t)$ is *PT* symmetric, i.e. $\hat{H}_{w,p}(t) = \hat{H}_{w,p}^{PT}(t)$ or $\hat{H}_{w,q}(t) = \hat{H}_{w,q}^{PT}(t)$, and a*nti-PT* symmetric weak value measurements are defined as those for which $\hat{H}_{w,p}(t)$ or $\hat{H}_{w,q}(t)$ is ant-$\mathcal{PT}$ symmetric, i.e., $\hat{H}_{w,p}^{PT}(t) = -\hat{H}_{w,p}(t)$ or $\hat{H}_{w,q}^{PT}(t) = -\hat{H}_{w,q}(t)$.

**Lemma 1.** $\hat{H}_{w,p}(t)$ and $\hat{H}_{w,q}(t)$ are $\mathcal{PT}$ symmetric if, and only if, $A_w(t)$ is even symmetric in time.

*Proof.* ($\Rightarrow$) If $\hat{H}_{w,p}(t)$ is $\mathcal{PT}$ symmetric, then $\gamma(t)A_w(-t)^*\hat{p} = \gamma(t)A_w(t)\hat{p}$. This implies that $A_w(-t)^* = A_w(t)$ in which case $A_w(t)$ is even symmetric in time. Similarly, for $\hat{H}_{w,q}(t)$. ($\Leftarrow$) If $A_w(t)$ is even symmetric in time, then $A_w(-t)^* = A_w(t)$ so that $\gamma(t)A_w(-t)^*\hat{p} = \gamma(t)A_w(t)\hat{p}$ in which case $\hat{H}_{w,p}(t) = \hat{H}_{w,p}^{PT}(t)$. Similarly, for $\hat{H}_{w,q}(t)$. ∎

**Theorem 2.** $\hat{H}_{w,p}(t)$ and $\hat{H}_{w,q}(t)$ are $\mathcal{PT}$ symmetric if, and only if, $Re\ A_w(t)$ is even symmetric in time and $Im(t)$ is odd symmetric in time.

*Proof.* ($\Rightarrow$) Let $\hat{H}_{w,p}(t)$ and $\hat{H}_{w,q}(t)$ be $\mathcal{PT}$ symmetric. Then $A_w(t)$ is even symmetric in time (Lemma 1), so that $A_w(-t)^* = A_w(t)$, in which case $Re\ A_w(-t) - i\ Im(-t) = Re\ A_w(t) + i\ Im(t)$. This condition only holds when $Re\ A_w(-t) = Re\ A_w(t)$, i.e., when $Re\ A_w(t)$ is even symmetric in time, and when $Im(-t) = -Im(t)$, i.e., when $Im(t)$ is odd symmetric in time. ($\Leftarrow$) If $Re\ A_w(t)$ is even symmetric in time and $Im(t)$ is odd symmetric in time, then $Re\ A_w(t) = Re\ A_w(-t)$ and $-Im(-t) = Im(t)$. In this case, $A_w(t) = Re\ A_w(t) + i\ Im(t) = Re\ A_w(-t) - i\ Im(-t) = A_w(-t)^*$ so that $A_w(t)$ is even symmetric in time. Consequently (Lemma 1), $\hat{H}_{w,p}(t)$ and $\hat{H}_{w,q}(t)$ are $\mathcal{PT}$ symmetric. ∎

**Lemma 3.** $\hat{H}_{w,p}(t)$ and $\hat{H}_{w,q}(t)$ are anti-$\mathcal{PT}$ symmetric if, and only if, $A_w(t)$ is odd symmetric in time.

*Proof.* ($\Rightarrow$) If $\hat{H}_{w,p}(t)$ is anti-$\mathcal{PT}$ symmetric, then $\gamma(t)A_w(-t)^*\hat{p} = -\gamma(t)A_w(t)\hat{p}$. This implies that $A_w(t) = -A_w(-t)^*$ in which case $A_w(t)$ is odd symmetric in time. Similarly, for $\hat{H}_{w,q}(t)$. ($\Leftarrow$) If $A_w(t)$ is odd symmetric in time, then $A_w(t) = -A_w(-t)^*$ so that $\gamma(t)A_w(t)\hat{p} = -\gamma(t)A_w(-t)^*\hat{p}$ or $\gamma(t)A_w(-t)^*\hat{p} = -\gamma(t)A_w(t)\hat{p}$ in which case $\hat{H}_{w,p}^{PT}(t) = -\hat{H}_{w,p}(t)$. Similarly, for $\hat{H}_{w,q}(t)$. ∎

**Theorem 4.** $\hat{H}_{w,p}(t)$ and $\hat{H}_{w,q}(t)$ are anti-$\mathcal{PT}$ symmetric if, and only if, $Re\ A_w(t)$ is odd symmetric in time and $Im(t)$ is even symmetric in time.

*Proof.* ($\Rightarrow$) Let $\hat{H}_{w,p}(t)$ and $\hat{H}_{w,q}(t)$ be anti-$\mathcal{PT}$ symmetric. Then $A_w(t)$ is odd symmetric in time (Lemma 3), so that $-A_w(-t)^* = A_w(t)$, in which case $-[Re\ A_w(-t) - i\ Im(-t)] = Re\ A_w(t) + i\ Im(t)$. This condition only holds when $-Re\ A_w(-t) = Re\ A_w(t)$ or $Re\ A_w(-t) = -Re\ A_w(t)$, i.e., when $Re\ A_w(t)$ is odd symmetric in time, and when $Im(-t) = Im(t)$, i.e., when $Im(t)$ is even symmetric in time. ($\Leftarrow$) If $Re\ A_w(t)$ is odd symmetric in time and $Im(t)$ is even symmetric in time, then $Re\ A_w(t) = -Re\ A_w(-t)$ and $Im(-t) = Im(t)$. In this case $A_w(t) = Re\ A_w(t) + i\ Im(t) = -Re\ A_w(-t) + i\ Im(-t) = -[Re\ A_w(t) - i\ Im(t)] = -A_w(-t)^*$. Consequently, (Lemma 3), $\hat{H}_{w,p}(t)$ and $\hat{H}_{w,q}(t)$ are anti-$\mathcal{PT}$ symmetric. ∎

Thus, the $\mathcal{PT}$ and anti-$\mathcal{PT}$ symmetry associated with weak value measurements of an observable depend strictly upon the even and odd time symmetries of $A_w(t)$, respectively (or equivalently upon the time symmetries of the real and imaginary parts of $A_w(t)$).

## 4 Consequences of $\mathcal{PT}$ and Anti-$\mathcal{PT}$ Symmetry for Weak Value Measurements

This section examines the weak position and weak momentum pointer translations generated by $\mathcal{PT}$ and anti-$\mathcal{PT}$ symmetric weak value measurements when the coupling strengths are proportional to a function that is even symmetric in time around a time $t_0$. In what follows, it is always assumed that: (i) the weakness conditions (eq. 3.5 in [3]) required for a weak value measurement are satisfied; (ii) the weak value has no vertical asymptotes; and (iii) the products of the coupling strengths and weak values are integrable.

**Theorem 5.** *PT symmetric measurements of weak values that are even symmetric in time around $t_0$ have weak pointer position and weak pointer momentum translations with vanishing imaginary parts.*

*Proof.* Since $A_w(t - t_0)$ is even symmetric in time around $t_0$, then $Im\ A_w(t - t_0)$ is odd symmetric in time around $t_0$ (Theorem 2). Therefore, the products $\gamma(t - t_0)\ Im\ A_w(t - t_0)$ and $\theta(t - t_0)\ Im\ A_w(t - t_0)$ must also be odd symmetric in time around $t_0$. The result follows from the application of the integration property of odd functions. ∎

As a simple illustration of this theorem, let the coupling strengths be $\gamma(t) = \gamma_0 \delta^{(\epsilon)}(t - t_0)$ and $\theta(t) = \theta_0 \delta^{(\epsilon)}(t - t_0)$, where $\delta^{(\epsilon)}(t - t_0)$ is the even symmetric in time (around $t_0$) function defined as

$$\delta^{(\epsilon)}(t - t_0) \equiv \begin{cases} \frac{1}{\epsilon} & t_0 - \frac{\epsilon}{2} \leq t \leq t_0 + \frac{\epsilon}{2} \\ 0 & |t - t_0| > \frac{\epsilon}{2} \end{cases}. \tag{12}$$

Assume the weak value is $A_w(t) = \cos 2\omega(t - t_0) + i \sin 2\omega(t - t_0)$ and observe that $A_w(t)$ is even symmetric in time and its real and imaginary parts are even and odd symmetric in time around $t_0$, respectively. Thus, $\widehat{H}_{w,p}(t)$ and $\widehat{H}_{w,q}(t)$ are PT symmetric operators (Lemma 1, Theorem 2) and the measurements of $A_w(t)$ are PT symmetric measurements. The associated weak pointer translations are

$$q_w = \frac{\gamma_0}{\epsilon} \int_{t_0 - \frac{\epsilon}{2}}^{t_0 + \frac{\epsilon}{2}} (\cos 2\omega(t - t_0) + i \sin 2\omega(t - t_0)) dt = \frac{\gamma_0}{\omega \epsilon} \sin \omega \epsilon \tag{13}$$

and

$$p_w = \frac{\theta_0}{\epsilon} \int_{t_0 - \frac{\epsilon}{2}}^{t_0 + \frac{\epsilon}{2}} (\cos 2\omega(t - t_0) + i \sin 2\omega(t - t_0)) dt = \frac{\theta_0}{\omega \epsilon} \sin \omega \epsilon. \tag{14}$$

As required by Theorem 5, the imaginary parts of the weak pointer translations vanish.

**Theorem 6.** *Anti-PT symmetric measurements of weak values that are odd symmetric in time around $t_0$ have weak pointer position and weak pointer momentum translations with vanishing real parts.*

*Proof.* Since $A_w(t - t_0)$ is odd symmetric in time around $t_0$, then $Re\ A_w(t - t_0)$ is odd symmetric in time around $t_0$ (Theorem 4). Therefore, the products $\gamma(t - t_0)\ Re\ A_w(t - t_0)$ and $\theta(t - t_0)\ Re\ A_w(t - t_0)$ must also be odd symmetric in time around $t_0$. The result follows from the application of the integration property of odd functions. ∎

As a simple illustration of this theorem, let the coupling strengths be as just defined - but now assume that $A_w(t) = \sin 2\omega(t - t_0) + i \cos 2\omega(t - t_0)$. By inspection, it is seen that $A_w(t)$ is odd symmetric in time around $t_0$ and it's real and imaginary parts are odd and even symmetric in time around $t_0$, respectively. In this case $\widehat{H}_{w,p}(t)$ and $\widehat{H}_{w,q}(t)$ are anti-PT symmetric operators (Lemma 3, Theorem 4) and the measurements of $A_w(t)$ are anti-PT symmetric measurements. The associated weak pointer translations are

$$q_w = \frac{\gamma_0}{\epsilon} \int_{t_0 - \frac{\epsilon}{2}}^{t_0 + \frac{\epsilon}{2}} (\sin 2\omega(t - t_0) + i \cos 2\omega(t - t_0)) dt = i \frac{\gamma_0}{\omega \epsilon} \sin \omega \epsilon \tag{15}$$

and

$$p_w = \frac{\theta_0}{\epsilon} \int_{t_0 - \frac{\epsilon}{2}}^{t_0 + \frac{\epsilon}{2}} (\sin 2\omega(t - t_0) + i \cos 2\omega(t - t_0)) dt = i \frac{\theta_0}{\omega \epsilon} \sin \omega \epsilon. \tag{16}$$

As required by Theorem 6, the real parts of the weak pointer translations vanish. (Observe that when $\omega \epsilon \ll 1$, then $\sin \omega \epsilon \approx \omega \epsilon$ and, as expected, the impulsive coupling results obtained using the Dirac delta "function" in place of $\delta^{(\epsilon)}(t - t_0)$ serve as good approximations for (13) - (16) ).

## 5 Application to Arbitrary Pointer Observables

Theorems 5 and 6 can be directly applied to the recently reported properties of weak value measurement pointers for the general case where $\langle q|\varphi\rangle$ and $A_w$ are complex valued and the interaction Hamiltonian is given by (3) with $\gamma(t) = \gamma_0 \delta(t-t_0)$ (recall that the Dirac delta "function" is time symmetric around $t_0$) [31,32]. There it was shown that the mean value and variance of an arbitrary pointer observable $M$ after such measurements are given by

$$\langle\Phi|\hat{M}|\Phi\rangle = \langle\varphi|\hat{M}|\varphi\rangle - i\left(\frac{\gamma_0}{\hbar}\right) Re\, A_w(t_0)\langle\varphi|[\hat{M},\hat{p}]|\varphi\rangle + \left(\frac{\gamma_0}{\hbar}\right) Im A_w(t_0)(\langle\varphi|\{\hat{M},\hat{p}\}|\varphi\rangle - 2\langle\varphi|\hat{M}|\varphi\rangle\langle\varphi|\hat{p}|\varphi\rangle) \quad (17)$$

and

$$\Delta_\Phi^2 M = \Delta_\varphi^2 M - i\left(\frac{\gamma_0}{\hbar}\right) Re\, A_w(t_0)\, F(\hat{M}) + \left(\frac{\gamma_0}{\hbar}\right) Im A_w(t_0)\, G(\hat{M}), \quad (18)$$

where $\Delta_\varphi^2 M$ and $\Delta_\Phi^2 M$ are the initial and final variances, respectively, $\{\hat{M},\hat{p}\} = \hat{M}\hat{p} + \hat{p}\hat{M}$,

$$\gamma_0 A_w(t_0) = \int_{-\infty}^{+\infty} \gamma_0 \delta(t-t_0) A_w(t) dt = p_w,$$

$$F(\hat{M}) \equiv \langle\varphi|[\hat{M}^2,\hat{p}]|\varphi\rangle - 2\langle\varphi|\hat{M}|\varphi\rangle\langle\varphi|[\hat{M},\hat{p}]|\varphi\rangle,$$

and

$$G(\hat{M}) \equiv \langle\varphi|\{\hat{M}^2,\hat{p}\}|\varphi\rangle - 2\langle\varphi|\hat{M}|\varphi\rangle\langle\varphi|\{\hat{M},\hat{p}\}|\varphi\rangle - 2\langle\varphi|\hat{p}|\varphi\rangle\left(\Delta_\varphi^2 M - \langle\varphi|\hat{M}|\varphi\rangle^2\right).$$

Now suppose the measurement is $\mathcal{PT}$ symmetric. From Theorem 5, $Im\, p_w = \gamma_0 Im\, A_w(t_0) = 0$ and (17) and (18) reduce to

$$\langle\Phi|\hat{M}|\Phi\rangle = \langle\varphi|\hat{M}|\varphi\rangle - i\left(\frac{\gamma_0}{\hbar}\right) Re\, A_w(t_0)\langle\varphi|[\hat{M},\hat{p}]|\varphi\rangle$$

and

$$\Delta_\Phi^2 M = \Delta_\varphi^2 M - i\left(\frac{\gamma_0}{\hbar}\right) Re\, A_w(t_0)\, F(\hat{M}).$$

Consequently, for $\mathcal{PT}$ symmetric measurements, the mean value and variance for an arbitrary pointer observable are only affected by the real part of the weak value being measured. For example, if $\hat{M} = \hat{q}, \hat{p}$, then $F(\hat{q}) = 0 = F(\hat{p})$ so that

$$\langle\Phi|\hat{q}|\Phi\rangle = \langle\varphi|\hat{q}|\varphi\rangle + \gamma_0 Re\, A_w(t_0), \qquad \Delta_\Phi^2 q = \Delta_\varphi^2 q$$

and

$$\langle\Phi|\hat{p}|\Phi\rangle = \langle\varphi|\hat{p}|\varphi\rangle, \qquad \Delta_\Phi^2 p = \Delta_\varphi^2 p.$$

Comparison of these results with (9) and (10) in [31] reveals that, since $\gamma_0 Im\, A_w(t_0) = 0$ for $\mathcal{PT}$ symmetric measurements, $\Delta_\varphi^2 q$ and $\Delta_\varphi^2 p$ have no affect upon $\langle\Phi|\hat{q}|\Phi\rangle$ and $\langle\Phi|\hat{p}|\Phi\rangle$ when $\langle q|\varphi\rangle$ and $A_w$ are complex valued.

If the measurement is anti-$\mathcal{PT}$ symmetric, then $Re\, p_w = \gamma_0 Re A_w(t_0) = 0$ (Theorem 6) and (12) and (13) reduce to

$$\langle\Phi|\hat{M}|\Phi\rangle = \langle\varphi|\hat{M}|\varphi\rangle + \left(\frac{\gamma_0}{\hbar}\right) Im A_w(t_0)(\langle\varphi|\{\hat{M},\hat{p}\}|\varphi\rangle - 2\langle\varphi|\hat{M}|\varphi\rangle\langle\varphi|\hat{p}|\varphi\rangle)$$

and

$$\Delta_\Phi^2 M = \Delta_\varphi^2 M + \left(\frac{\gamma_0}{\hbar}\right) Im A_w(t_0)\, G(\hat{M}).$$

Thus, for anti-$\mathcal{PT}$ symmetric measurements, the mean value and variance for any pointer observable are affected only by the imaginary part of the weak value being measured. When $\hat{M} = \hat{q}, \hat{p}$, then $G(\hat{q}) = \frac{2m}{3}\dot{q}_3$, $G(\hat{p}) = 2p_3$,

$$\langle\varphi|\{\hat{q},\hat{p}\}|\varphi\rangle - 2\langle\varphi|\hat{q}|\varphi\rangle\langle\varphi|\hat{p}|\varphi\rangle = m(\Delta_\varphi^2 q),$$

and

$$\langle\varphi|\{\hat{p},\hat{p}\}|\varphi\rangle - 2\langle\varphi|\hat{p}|\varphi\rangle\langle\varphi|\hat{p}|\varphi\rangle = \Delta_\varphi^2 p,$$

[24], in which case

$$\langle\Phi|\hat{q}|\Phi\rangle = \langle\varphi|\hat{q}|\varphi\rangle + m\left(\frac{\gamma_0}{\hbar}\right)ImA_w(t_0)(\Delta_\varphi^2 \dot{q}), \quad \Delta_\Phi^2 q = \Delta_\varphi^2 q + \frac{2m}{3}\left(\frac{\gamma_0}{\hbar}\right)ImA_w(t_0)\dot{q}_3$$

and

$$\langle\Phi|\hat{p}|\Phi\rangle = \langle\varphi|\hat{p}|\varphi\rangle + 2\left(\frac{\gamma_0}{\hbar}\right)ImA_w(t_0)(\Delta_\varphi^2 p), \quad \Delta_\Phi^2 p = \Delta_\varphi^2 p + 2\left(\frac{\gamma_0}{\hbar}\right)ImA_w(t_0)\, p_3.$$

Here, $m$ is the mass of the pointer, $x_3 \equiv \langle\varphi|(\hat{x} - \langle\varphi|\hat{x}|\varphi\rangle)^3|\varphi\rangle$, $x = q, p$, and $\dot{q}_3$ and $(\dot{q}\Delta_\varphi^2)$ are rates of change as $t$ approaches $t_0$. Comparison of these results with (5) in [31] shows that, since $\gamma_0 ReA_w(t_0) = 0$ for anti-*PT* symmetric measurements, the pointer position does not respond to $\gamma_0 ReA_w(t_0)$.

## 6 Closing Remarks

This paper has examined the parity *P* and time reversal *T* invariance of the non-Hermitian Hamiltonians naturally appearing in the von Neumann interaction operator model for measurements of time dependent weak values. It was shown that, under the fairly mild restrictions of integrability and time symmetric coupling strengths, these invariances depend upon the time symmetry of the weak value being measured and yield either real or imaginary valued pointer translations. In particular, *PT* (anti-*PT*) symmetric measurements require the weak value to be even (odd) symmetric in time and produce only real (imaginary) valued translations. Thus, *PT* (anti-*PT*) symmetric measurements have predictable consequences for arbitrary pointer observables in the sense that the associated [32] pointer translation and variance are only affected by the real (imaginary) part of the weak value being measured.

We note that the two-time formulation of QM may be fertile ground for exploring *PT* symmetry via time-dependent weak values, so they could provide potential prediction of symmetries that might be observable and thus, provides means to explore differences between the various two-time interpretations that are distinct from the Aharonov version. Also, we note the speculative interpretation, for the moment, of *T*: *This interval, which can be loosely interpreted by what we mean by the terminology "Present" to be an aspect of physics rather a psychological manifestation of the mind. So, we might speculate this is another reason the two-time formulation being physics rather than philosophy. The Aharonov interpretation of quantum mechanics might provide a physical reason for the existence of the physical manifestations of the notion of the present by requiring the existence of weak energy. Weak energy is an aspect of this manifestation by the composition of the two directions of evolution interacting to create the interval T, which we propose be interpreted as the "present interval" rather than as a moment, which is discussed in detail Section 2.*

In closing, it is noted that investigation into the possible relationships between *PT* (anti-*PT*) symmetric weak value measurements and the Dressel and Jordan interpretation of the role of the imaginary part of the weak value [33] might prove interesting and useful to be explored further in the future.

### Acknowledgment

This work was supported in part by a grant from the Naval Surface Warfare Center, Dahlgren Division, In-House Laboratory Independent Research program sponsored by the Office of Naval Research.

### References

[1] Aharonov, Y., Albert, D., Casher, A., Vaidman, L. : Novel properties of preselected and postselected ensembles. In: Greenberger,D. (ed.) New techniques and ideas in quantum measurement theory, 417-421. New York Academy of Sciences, New York (1986).

[2] Aharonov, Y., Albert, D., Vaidman, L.: How the result of a measurement of a component of the spin of a spin-1/2 particle can turn out to be 100. Phys. Rev. Lett. 60, 1351-1354 (1988). https://doi.org/10.1103/PhysRevLett.60.1351.

[3] Aharonov, Y., Vaidman, L.: Properties of a quantum system during the time interval between two measurements. Phys. Rev. A 41, 11-20 (1990). https://doi.org/10.1103/PhysRevA.41.11.

[4] Hosten, O., Kwiat, P.: Observation of the spin Hall effect of light via weak measurements. Science 319, 787-790 (2008). https://doi.org/10.1126/science.1152697.

[5] Piacentini, F., Avella, A., Levi, M.P., Lussana, R., Villa, F., Tosi, A., Zappa, F., Gramenga, M., Brida, G., Degiovanni, I.P., Genovese, M.: Experiment investigating the connection between weak values and contextuality. Phys. Rev. Lett. 116, 180401 (5pp) (2016). https://doi.org/10.1103/PhysRevLett.116.180401.

[6] Spence, S., Parks, A.: Experimental evidence for retro-causation in quantum mechanics using weak values. Quantum Stud. Math. Found. 4, 1-6 (2017). https://doi.org/10.1007/s40509-016-0082-x.

[7] Aharonov, Y., Botero, A., Popescu, S., Reznik, B., Tollaksen, J.: Revisiting Hardy's paradox: counterfactual statements, real measurements, entanglement and weak values. Physics Letters A 301, 130-138 (2002). https://doi.org/10.1016/S0375-9601(02)00986-6.

[8] Parks, A., Spence, S., Farinholt, J.: A note concerning the modular valued von Neumann interaction operator. Quantum Stud. Math. Found. 6, 101-105 (2019). https://doi.org/10.1007/s40509-018-0167-9.

[9] Brunner, N., Simon, C.: Measuring small longitudinal phase shifts: weak measurements or standard interferometry?. Phys. Rev. Lett. 105, 010405 (4pp) (2010). https://doi.org/10.1103/PhysRevLett.105.010405

[10] Howland, G., Lum, D., Howell, J.: Compressive wave front sensing with weak values. Optics Express 22, 18870-18880 (2014). https://doi.org/10.1364/OE.22.018870.

[11] Parks, A., Spence, S.: Comparative weak value amplification as an approach to estimating the value of small quantum mechanical interactions. Metrol. Meas, Syst. 23, 393-401 (2016). https://doi.org/10.1515/mms-2016-0035.

[12] Bender, C., Boettcher, S.: Real Spectra in Non-Hermitian Hamiltonians Having PT Symmetry. Phys. Rev. Lett. 80, 5243-5246 (1998). https://doi.org/10.1103/PhysRevLett.80.5243.

[13] Bender, C., Boettcher, S., Meisinger, P.: PT-symmetric quantum mechanics. J. Math. Phys. 40, 2201-2229 (1999). https://doi.org/10.1063/1.532860.

[14] Longhi, S.: Optical realization of relativistic non-Hermitian quantum mechanics. Phys. Rev. Lett. 105, 013903 (4pp) (2010). https://doi.org/10.1103/PhysRevLett.105.013903.

[15] Chtchelkatchev, N., Golubov, A., Baturina, T., Vinokur, V.: Stimulation of the fluctuation superconductivity by PT symmetry. Phys. Rev. Lett. 109, 150405 (5pp) (2012). https://doi.org/10.1103/PhysRevLett.109.150405.

[16] Hang, C., Huang, G., Knontop, V.: PT symmetry with a system of three-level atoms. Phys. Rev. Lett. 110, 083604 (5pp) (2013). https://doi.org/10.1103/PhysRevLett.110.083604.

[17] Wu, Y., Liu, W., Geng, J., Song, X., Ye, X., Duan, C., Rong, X., Du, J.: Observation of parity-time symmetry breaking in a single-spin system. Science 364, 878-880 (2019). https://doi.org/10.1126/science.aaw8205.

[18] Benisty, H., Degiron, A., Lupu, A., De Lustrac, A., Chenais, S., Forget, S., Besbes, M., Barbillon, G., Bruyant, A., Blaize, S., Lerondel, G.: Implementation of PT symmetric devices using plasmonics: principle and applications. Optics Express 19, 18004-18019 (2011).

[19] Feng, L., Xu, Y., Fegadolli, W., Lu, M., Oliveira, J., Almeida, V., Chen, Y., Scherer, A.: Experimental demonstration of a unidirectional reflectionless parity-time metamaterial at optical frequencies. Nature Materials 12, 108-113 (2012). https://doi.org/10.1038/NMAT3495.

[20] Pati, K., Singh, U., Sinha, U.: Measuring non-Hermitian operators via weak values. Phys. Rev. A 92, 052120 (8pp) (2015). https://doi.org/10.1103/PhysRevA.92.052120.

[21] Aharonov, Y. and Vaidman, L., 2008. The two-state vector formalism: an updated review. *Time in quantum mechanics*, pp.399-447.

[22] Aharonov, Y., Cohen, E. and Landsberger, T., 2017. The two-time interpretation and macroscopic time-reversibility. *Entropy*, *19*(3), p.111.

[23] Aharonov, Y., Cohen, E., Waegell, M. and Elitzur, A.C., 2018. The weak reality that makes quantum phenomena more natural: novel insights and experiments. *Entropy*, *20*(11), p.854.

[24] Robertson, K., 2017. Can the two-time interpretation of quantum mechanics solve the measurement problem?. *Studies in History and Philosophy of Science Part B: Studies in History and Philosophy of Modern Physics*, *58*, pp.54-62.

[25] Parks, A.D., 2008. Time-dependent weak values and their intrinsic phases of evolution. *Journal of Physics A: Mathematical and Theoretical*, *41*(33), p.335305.

[26] Parks, A.D., 2000. The geometry and significance of weak energy. *Journal of Physics A: Mathematical and General*, *33*(13), p.2555.
[27] Pollak, E. and Miret-Artés, S., 2019. Uncertainty relations for time-averaged weak values. *Physical Review A*, *99*(1), p.012108.
[28] Davies, P.C.W., 2009. Time-dependent quantum weak values: decay law for postselected states. *Physical Review A*, *79*(3), p.032103.
[29] Mori, Y. and Tsutsui, I., 2021. Weak value amplification and the lifetime of decaying particle. *arXiv preprint arXiv:2105.12349*.
[30] Hassanpour, N.: Topics in PT-symmetric Quantum Mechanics and Classical Systems. Arts & Sciences Electronic Theses and Dissertations 1625, p.2 (2018). https://openscholarship.wustl.edu/art_sci_etds/1625.
[31] Parks, A., Cullin, D., Stoudt, D.: Observation and measurement of an optical Aharonov-Albert-Vaidman effect. Proc. R. Soc. Lond. A 454, 2997-3008 (1998).
[31] Jozsa, R.: Complex weak values in quantum measurement. Phys. Rev. A 76, 044103 (2007). https://doi.org/10.1103/PhysRevA.76.044103.
[32] Parks, A., Gray, J.: Variance control in weak-value measurement pointers. Phys. Rev. A 84, 012116 (4pp) (2011). https://doi.org/10.1103/PhysRevA.84.012116.
[33] Dressel, J., Jordan, A.: Significance of the imaginary part of the weak value. Phys. Rev. A 85, 012107 (13pp) (2012). https://doi.org/10.1103/PhysRevA.85.012107.